\documentclass[copyright]{eptcs}
\providecommand{\event}{J.~Dassow, G.~Pighizzini, B.~Truthe (Eds.): 11th International Workshop\\
on Descriptional Complexity of Formal Systems (DCFS 2009)} 
\providecommand{\volume}{3}
\providecommand{\anno}{2009}
\providecommand{\firstpage}{53} 
\providecommand{\eid}{5}
\usepackage{breakurl}        

\input{dcfs.tex}

\usepackage[utf8]{inputenc}
\usepackage{xspace}
\usepackage{amsmath}
\usepackage{amssymb}
\usepackage{listings}
\usepackage{url}
\usepackage{vaucanson-g}

\newcommand{\dfa}{DFA\xspace}

\newcommand{\dfas}{DFAs\xspace}
\newcommand{\nfa}{NFA\xspace}
\newcommand{\nfas}{NFAs\xspace}

\newcommand{\icdfa}{ICDFA\xspace}

\newcommand{\ACI}{$ACI$\xspace}

\newcommand{\RE}{RE\xspace}

\newcommand{\re}{r.\,e.\xspace}

\newcommand{\const}{\operatorname{\varepsilon}}

\newcommand{\derivative}[1]{\mbox{${#1}^{-1}$}}

\newcommand{\ie}{i.\,e.}
\newcommand{\cf}{cf.\xspace}

\newcommand{\hop}{\textbf{Hop}\xspace}
\newcommand{\equivp}{\textbf{equivP}\xspace}
\newcommand{\equivuf}{\textbf{equivUF}\xspace}
\newcommand{\hk}{\textbf{HK}\xspace}
\newcommand{\hkn}{\textbf{HKn}\xspace}
\newcommand{\hki}{\textbf{HKi}\xspace}
\newcommand{\hke}{\textbf{HKe}\xspace}
\newcommand{\hks}{\textbf{HKs}\xspace}
\newcommand{\am}{\textbf{AM}\xspace}
\newcommand{\brz}{\textbf{Brz}\xspace}
\renewcommand{\star}{*}

\begin{document}
\title{Testing the Equivalence of Regular Languages\,%
\thanks{This work was partially funded by Funda\c{c}\~ao para a
  Ci\^encia e Tecnologia (FCT) and Program POSI, and by project ASA
  (PTDC/MAT/65481/2006).}}
\def\titlerunning{Testing the Equivalence of Regular Languages}

\author{%
\makebox[0cm][c]{%
Marco Almeida\,\thanks{Marco Almeida is funded by FCT grant SFRH/BD/27726/2006.}}%
\hspace{40mm}%
\makebox[0cm][c]{%
Nelma Moreira}%
\hspace{40mm}%
\makebox[0cm][c]{%
Rog\'erio Reis}
\institute{DCC-FC \ \& LIACC -- Universidade do Porto \\
   R. do Campo Alegre 1021/1055 -- 4169-007 Porto -- Portugal}
\email{\makebox[0cm][c]{mfa@ncc.up.pt}\hspace{40mm}%
       \makebox[0cm][c]{nam@ncc.up.pt}\hspace{40mm}%
       \makebox[0cm][c]{rvr@ncc.up.pt}}
}

\def\authorrunning{M.~Almeida, N.~Moreira, R.~Reis}

\maketitle

\begin{abstract}
  The minimal deterministic finite automaton is generally used to
  determine regular languages equality. Antimirov and Mosses proposed
  a rewrite system for deciding regular expressions equivalence of
  which Almeida {\em et al.} presented an improved variant. Hopcroft
  and Karp proposed an almost linear algorithm for testing the
  equivalence of two deterministic finite automata that avoids
  minimisation.
  In this paper we improve the best-case running time, present an
  extension of this algorithm to non-deterministic finite automaton,
  and establish a relationship between this algorithm and the one
  proposed in Almeida {\em et al.} We also present some experimental
  comparative results. All these algorithms are closely related with
  the recent coalgebraic approach to automata proposed by Rutten.
\end{abstract}

\section{Introduction}

The uniqueness of the minimal deterministic finite automaton for each
regular language is in general used for determining regular languages
equality. Whether the languages are represented by deterministic
finite automata (\dfa), non deterministic finite automata (\nfa), or
regular expressions (\re), the usual procedure uses the equivalent
minimal \dfa{} to decide equivalence. The best known algorithm, in
terms of worst-case analysis, for \dfa minimisation is loglinear
\cite{hopcroft71_c}, and the equivalence problem is PSPACE-complete
for both \nfa and \re Based on the algebraic properties of regular
expressions, Antimirov and Mosses proposed a terminating and complete
rewrite system for deciding their
equivalence~\cite{antimirov94:_rewrit_exten_regul_expres}. In a paper
about testing the equivalence of regular expressions, Almeida {\em et
  al.}~\cite{almeida08_c:_antim_and_mosses_rewrit_system_revis}
presented an improved variant of this rewrite system.  As suggested by
Antimirov and Mosses, and corroborated by further experimental
results, a better average-case performance may be obtained.

Hopcroft and Karp~\cite{hopcroft71_c:_linear_algor_for_testin_equiv}
presented, in 1971, an almost linear algorithm for testing the
equivalence of two \dfa{}s that avoids their minimisation. Considering
the merge of the two \dfas as a single one, the algorithm computes the
finest right-invariant relation which identifies the initial
states. The state equivalence relation that determines the minimal
\dfa{} is the coarsest relation in that condition.

We present some variants of Hopcroft and Karp's algorithm (\hk{})
(Section~\ref{sec:dfa}), and establish a relationship with the one
proposed in Almeida \emph{et al.}
(Section~\ref{sec:relationship_with_am}).  In particular, we extend
\hk{} algorithm to \nfa{}s and present some experimental comparative
results (Section~\ref{sec:experimental}).

All these algorithms are also closely related with the recent
coalgebraic approach to automata developed by
Rutten~\cite{rutten03:_behav}, where the notion of \emph{bisimulation}
corresponds to a right-invariance. Two automata are bisimilar  if there
exists a bisimulation between them. For deterministic (finite)
automata, the \emph{coinduction proof principle} is effective for
equivalence, \ie, two automata are bisimilar if and only if they are
equivalent. Both Hopcroft and Karp algorithm and Antimirov and Mosses
method can be seen as instances of this more general approach
(\cf Corollary~\ref{cor:1}). This means that these  methods may be
easily extended to other Kleene Algebras,
namely the ones that model program properties, and that have been
successfully applied in formal program verification~\cite{kozen08}.

\section{Preliminaries}\label{sec:pre}
We recall here the basic definitions needed throughout the paper.
For further details we refer the reader to the works of Hopcroft
\emph{et al.}  \cite{hopcroft00_c:_introd_autom_theor_languag_comput}
and Kozen \cite{kozen97:_autom_comput}.

 A \emph{regular expression} (\re)
$\alpha$ over an alphabet $\Sigma$ represents a (regular) language
$L(\alpha)\subseteq \Sigma^\star$ and is inductively defined by:
$\emptyset$ is a \re and $L(\emptyset)=\emptyset$; $\epsilon$ is a \re
and $L(\epsilon)=\{\epsilon\}$; $a\in \Sigma$ is a \re and
$L(a)=\{a\}$; if $\alpha$ and $\beta$ are \re, $(\alpha_1 +
\alpha_2)$, $(\alpha_1 \alpha_2)$ and $(\alpha_1)^\star$ are \re,
respectively with $L((\alpha_1 + \alpha_2))=L(\alpha_1)\cup
L(\alpha_2)$, $L((\alpha_1\alpha_2))=L(\alpha_1)L(\alpha_2)$ and
$L((\alpha_1)^\star)=L(\alpha_1)^\star$. We define $\const(\alpha)=1$
(resp. $\const(\alpha)=0$) if $\epsilon \in L(\alpha)$
(resp. $\epsilon \notin L(\alpha)$). Two \re $\alpha$ and $\beta$ are
\emph{equivalent}, and we write $\alpha\sim \beta$, if
$L(\alpha)=L(\beta)$.
The algebraic structure $(\RE,+,\cdot,\emptyset,\epsilon)$, where $\RE$
denotes the set of \re over $\Sigma$, constitutes an idempotent
semiring, and, with the unary operator $\star$, a \emph{Kleene
  algebra}. There are several well-known complete axiomatizations of
Kleene 
 algebras. 
Let \ACI denote the associativity, commutativity and idempotence
of~$+$.

A \emph{nondeterministic finite automaton} (\nfa) $A$ is a tuple
$(Q,\Sigma,\delta,I,F)$ where $Q$ is a finite set of states, $\Sigma$ is
the alphabet, $\delta \subseteq Q \times \Sigma \times Q$ the
transition relation, $I\subseteq Q$ the set of initial states, and $F
\subseteq Q$ the set of final states.  An \nfa is \emph{deterministic}
(\dfa) if for each pair $(q,a) \in Q \times \Sigma$ there exists at
most one $q'$ such that $(q,a,q') \in \delta$. The size of a \nfa is $|Q|$. For $s\in Q$ and $a\in \Sigma$, we denote by
$\delta(q,a)=\{p\mid(q,a,p)\in \delta\}$, and we can extend this
notation to $x\in \Sigma^\star$, and to $R\subseteq Q$. For a \dfa, we
consider $\delta : Q \times \Sigma^\star\rightarrow Q$.  The
\emph{language} accepted by $A$ is $L(A)=\{x\in \Sigma^\star\mid
\delta(I,x)\cap F\not= \emptyset\}$. Two \nfas $A$ and $B$ are
\emph{equivalent}, denoted by $A\sim B$ if they accept the same
language. Given an \nfa $A=(Q_N,\Sigma,\delta_N,I,F_N)$, we can use
the \emph{powerset construction} to obtain a \dfa{} $D=(Q_D, \Sigma,
\delta_D, q_0, F_D)$ equivalent to $A$, where $ Q_D = 2^{Q_N}$, $q_0 =
I$, for all $R\in Q_D$, $R\in F_D$ if and only $R\cap F_N \neq
\emptyset$, and for all $a\in \Sigma$, $\delta_D(R, a) = \bigcup_{q
  \in R} \delta_N(q, a)$. This construction can be optimised by
omitting states $R\in Q_D$ that are unreachable from the initial~state.

Given a finite automaton $(Q,\Sigma,\delta,q_0,F)$, let $\const(q)=1$
if $q\in F$ and $\const(q)=0$ otherwise. We call a set of states
$R\subseteq Q$ \emph{homogeneous} if for every \hbox{$p, q \in R$},
\hbox{$\const(p) = \const(q)$}. A \dfa is \emph{minimal} if there is no
equivalent \dfa with fewer states. Two states $q_1, q_2 \in Q$ are
said to be \emph{equivalent}, denoted $q_1 \sim q_2$, if for every $w
\in \Sigma^\star$, $\const(\delta(q_1, w))=\const(\delta(q_2,
w))$. Minimal \dfas are unique up to isomorphism. Given a \dfa~$D$,
the equivalent minimal \dfa $D/_{\!\!\sim}$ is called the
\emph{quotient automaton} of $D$ by the equivalence relation  $\sim$.  
The state equivalence relation~$\sim$,
is a special case of a right-invariant equivalence relation
w.\,r.\,t. $D$, \ie, a relation $\equiv\; \subseteq Q\times Q$ such that
all classes of $\equiv$ are homogeneous, and for any $p, q\in Q$,
$a\in \Sigma$ if $p\equiv q$, then $\delta(p,a)/_{\!\!\equiv} =
\delta(q,a)/_{\!\!\equiv}$, where for any set $S$,
$S/_{\!\!\equiv}=\{[s]\mid s\in S\}$.  Finally, we recall that every
equivalence relation $\equiv$ over a set $S$ is efficiently
represented by the partition of $S$ given by $S/_{\!\!\equiv}$.  Given two
equivalence relations over a set $S$, $\equiv_R$ and $\equiv_T$, we
say that $\equiv_R$ is \emph{finer} then $\equiv_T$ (and $\equiv_T$
\emph{coarser} then $\equiv_R$) if and only if $\equiv_R\subseteq
\equiv_T$.

\section{Testing finite automata equivalence}
\label{sec:dfa}

The classical approach to the comparison of \dfas relies on the
construction of the minimal equivalent \dfa. The best known algorithm
for this procedure runs in $O(k n \log n)$ time \cite{hopcroft71_c},
for a \dfa with $n$ states over an alphabet of $k$ symbols.
Hopcroft and Karp~\cite{hopcroft71_c:_linear_algor_for_testin_equiv}
proposed an algorithm for testing the equivalence of two \dfas that
makes use of an almost $O(n)$ set merging method.

\subsection{The original Hopcroft and Karp algorithm}
\label{subsec:original}
Let $A = (Q_1, \Sigma, p_0, \delta_1, F_1)$ and $B = (Q_2, \Sigma,
q_0, \delta_2, F_2)$ be two \dfas, with $|Q_1|=n$, $|Q_2|=m$, and such
that $Q_1$ and $Q_2$ are 
disjoint.
In order to simplify notation, we assume $Q = Q_1 \cup Q_2$, $F = F_1
\cup F_2$, and $\delta(p, a) = \delta_i(p, a)$ for $p \in Q_i$.
We begin by presenting the original algorithm by Hopcroft and Karp
\cite{aho74:_desig_analy_comput_algor} for testing the equivalence of
two \dfas as Algorithm~\ref{alg:hk0}.

If $A$ and $B$ are equivalent \dfas, the algorithm computes the finest
right-invariant equivalence relation over $Q$ that identifies the
initial states, $p_0$ and~$q_0$. The associated set partition is built
using the UNION-FIND method. This algorithm assumes disjoint sets and
defines the three functions which follow.
\begin{itemize}
\item MAKE($i$): creates a new set (singleton) for one element $i$
  (the identifier);
\item FIND($i$): returns the identifier $S_i$ of the set which contains $i$;
\item UNION($i, j, k$): combines the sets identified by $i$ and $j$ in
  a new set \hbox{$S_k = S_i \cup S_j$}; $S_i$ and $S_j$ are destroyed.
\end{itemize}

It is clear that, disregarding the set operations, the worst-case time
of the algorithm is $O(k(n+m))$, where
$k=|\Sigma|$. 
An arbitrary sequence of $i$ MAKE, UNION, and FIND operations, $j$ of
which are MAKE operations in order to create the required sets, can be
performed in worst-case time $O(i \alpha(j))$, where $\alpha(j)$ is
related to a functional inverse of the Ackermann function, and, as
such, grows \emph{very} slowly. In fact, for every \emph{practical}
values of $j$~(up to $2^{2^{2^{16}}}$),~$\alpha(j) \leq 4$. 

\begin{quote}\begin{lstlisting}[caption={The original \hk{} algorithm.},
captionpos=b,numbers=left,label=alg:hk0]
def HK($A,B$):
   for $q \in Q$: MAKE($q$)
   S = $\emptyset$
   UNION($p_0,q_0,q_0$); PUSH(S,$(p_0,q_0)$)
   while $(p,q)$ = POP(S):
      for $a \in \Sigma$:
         $p^\prime$ = FIND($\delta(p, a)$)
         $q^\prime$ = FIND($\delta(q, a)$)
         if $p^\prime \neq q^\prime$:
            UNION($p^\prime$,$q^\prime$,$q^\prime$)
            PUSH(S,$(p^\prime, q^\prime)$)
   if $\forall S_i \forall p, q \in S_i \quad \const(p) = \const(q): $ return True
   else: return False
\end{lstlisting}
\end{quote}

 When
applied to Algorithm \ref{alg:hk0}, this set union algorithm allows
for a worst-case time complexity of $O(k(n+m)+3i
\alpha(j))=O(k(n+m)+3(n+m) \alpha(n+m))$. Considering $\alpha(n+m)$
constant, the asymptotic running-time of the algorithm is
$O(k(n+m))$. The correctness of this algorithm is proved in
Section~\ref{sec:relationship_with_am}, Theorem~\ref{th:dfa_equiv}.

\subsection{Improved best-case running time}
\label{subsec:refutation}

By altering the FIND function in order to create the set being looked
for if it does not exist, \ie, whenever FIND($i$) fails, MAKE($i$) is
called and the set $S_i = \{i\}$ is created, we may add a
\emph{refutation} procedure earlier in the algorithm. This allows the
algorithm to return as soon as it finds a pair of states such that one
is final and the other is
not.
This alteration to the FIND procedure avoids the initialization of
$m+n$ sets which may never actually be used. These modifications to
Algorithm \ref{alg:hk0} are presented in Algorithm \ref{alg:hk1}.

Although it does not change the worst-case complexity, the best-case
analysis is considerably better, as it goes from $\Omega(k(n+m))$ to
$\Omega(1)$. Not only it is possible to distinguish the automata~by
the first pair of states, but it is also possible to avoid the linear
check in the lines 12--13.
The observed asymptotic behaviour of minimality of initially connected
\dfas (\icdfa{}s) \cite{almeida07_c:_enumer_gener_strin_autom_repres},
suggests that, when dealing with random \dfas, the probability of
having two equivalent automata is very low, and a refutation method
will be very useful (see Section \ref{sec:experimental}).

\begin{lemma}
\label{lem:refutation_is_ok} 
In line~5 of Algorithm~\ref{alg:hk0}, all the sets $S_i$ are
homogeneous if and only if all the pairs of states $(p, q)$ pushed
into the stack are such that $\const(p) = \const(q)$.
\begin{proof}
  \emph{
  Let us proceed by induction on the number $l$ of times line~5 is
  executed. If $l=1$, it is trivial. Suppose that lemma is true for
  the $l^{th}$ time the algorithm executes line~5. If for all $a\in
  \Sigma$, the condition in line~9 is false, for the $(l+1)^{th}$ time
  the homogeneous character of the sets remains unaltered. Otherwise,
  it is clear that in lines 10--11, $S_{p'}\cup S_{q'}$ is homogeneous
  if and only if $\const(p') = \const(q')$. Thus the lemma is true.}
\end{proof}

\end{lemma}
\begin{quote}
\begin{lstlisting}[caption={\hk{} algorithm with an early refutation step~(\hki).},captionpos=b,numbers=left,label=alg:hk1]
def HKi($A,B$):
   MAKE($p_0$); MAKE($q_0$)
   S = $\emptyset$
   UNION($p_0,q_0,q_0$); PUSH(S,$(p_0, q_0)$)
   while $(p,q)$ = POP(S):
      if $\const(p) \neq \const(q)$: return False
      for $a \in \Sigma$:
         $p^\prime$ = FIND($\delta(p,a)$)
         $q^\prime$ = FIND($\delta(q,a)$)
         if $p^\prime \neq q^\prime:$
            UNION($p^\prime, q^\prime, q^\prime$)
            PUSH(S,$(p^\prime, q^\prime)$)
   return True
\end{lstlisting}
\end{quote}

\begin{theorem}
  \label{th:refutation_is_ok}
  Algorithms~\ref{alg:hk0} (\hk) and~\ref{alg:hk1} (\hki) are
  equivalent.
  \begin{proof}
  \emph{
      By Lemma~\ref{lem:refutation_is_ok}, if there is a pair of
      states $(p, q)$ pushed into the stack such that $\const(p) \neq
      \const(q)$, then the algorithm can terminate and return
      \emph{False}. That is exactly what Algorithm~\ref{alg:hk1}
      does.}
  \end{proof}
\end{theorem}

\subsection{Testing \nfa equivalence}
\label{sec:nfa}

It is possible to extend Algorithm \ref{alg:hk1} to test the
equivalence of \nfas. The basic idea is to embed the powerset
construction into the algorithm, although this must be done with some
caution. 
Because of space limitations, we will only sketch this extension. 
We call this algorithm \hke.

Let $N_1 = (Q_1, \Sigma, \delta_1, I_1, F_1)$ and $N_2 = (Q_2, \Sigma,
\delta_2, I_2, F_2)$ be two NFAs. We assume that $Q_1$ and $Q_2$
disjoint, and, we make $Q_N = Q_1 \cup Q_2$, $F_N = F_1 \cup F_2$, and
$\delta_N(p, a) = \delta_i(p, a)$ for $p \in Q_i$.
Consider Algorithm~\ref{alg:hk1} with the following data: $q_0=I_1$,
$p_0=I_2$, and for $p\in 2^{Q_N}$, $\delta(p,a)= \bigcup_{q \in p}
\delta_N(q,a)$ and $\const(p) = 1$ if and only if $\exists q\in p :
\const(q) = 1$. Notice that when dealing with \nfas it is essential to
use the idea described in Subsection \ref{subsec:refutation} and to
adjust the FIND operation so that FIND($i$) creates the set $S_i$ if
it does not exist. This way we avoid calling MAKE for each of the
$2^{|Q_N|}$ sets, which would lead directly to the worst-case of the
powerset construction.
\begin{theorem}
  Algorithm \ref{alg:hk1} can be applied to NFAs by embedding the
  powerset construction method.
\end{theorem}

As any \dfa is a particular case of an \nfa, all the experimental
results presented on Section \ref{sec:experimental} use Algorithm
\hke, whether the finite automata being tested are deterministic or
not.

\section{Relationship with Antimirov and Mosses' method}
\label{sec:relationship_with_am}

\subsection{Antimirov and Mosses' algorithm}
\label{sec:am}

The \emph{derivative} \cite{brzozowski64:_deriv_of_regul_expres} of a
\re $\alpha$ with respect to a \emph{symbol} $a \in \Sigma$, denoted
$\derivative{a}(\alpha)$, is defined recursively on the structure of
$\alpha$ as follows: {
\begin{alignat*}{2}
  \derivative{a}(\emptyset) &= \emptyset; &\qquad \derivative{a}(\alpha+\beta) &= \derivative{a}(\alpha)+\derivative{a}(\beta); \\
  \derivative{a}(\epsilon) &= \emptyset; &\qquad \derivative{a}(\alpha\beta) &= \derivative{a}(\alpha)\beta+\const(\alpha)\derivative{a}(\beta);  \\
  \derivative{a}(b) &= 
  \begin{cases}
    \epsilon,&  \text{if $b=a$}; \\
    \emptyset,&  \text{otherwise}; \\ 
  \end{cases} &\qquad \derivative{a}(\alpha^\star) &= \derivative{a}(\alpha)\alpha^\star. 
\end{alignat*}
} 

This notion can be trivially extended to words, and considering \re
modulo the \ACI axioms, Brzo\-zow\-ski
\cite{brzozowski64:_deriv_of_regul_expres} proved that, the set of
derivatives of a \re~$\alpha$,~${\cal D}(\alpha)$, is finite. This
result leads to the definition of \emph{Brzozowski's automaton} which
is equivalent to a given \re $\alpha$: $D_\alpha = ({\cal
  D}(\alpha),\Sigma, \delta_\alpha, \alpha, F_\alpha)$ where
\hbox{$F_\alpha=\{d\in {\cal D}(\alpha)\mid \const(d)=\epsilon\}$}, and
$\delta_\alpha(d,a)=\derivative{a}(d)$, for all $d\in {\cal
  D}(\alpha)$, $a\in \Sigma$.

Antimirov and Mosses~\cite{antimirov94:_rewrit_exten_regul_expres}
proposed a rewrite system for deciding the equivalence of two extended \re 
(with intersection), based on a complete axiomatization. This is a
refutation method such that testing the equivalence of two~\re
corresponds to an iterated process of testing the equivalence of their
derivatives. In the process, a Brzozowski's automaton is computed for
each \re Not considering extended \re, Algorithm~\ref{alg:am} is a
version of \am's method, which was, essentially, the one proposed by
Almeida \emph{et al.}
\cite{almeida08_c:_antim_and_mosses_rewrit_system_revis}.
\vspace{-0.3cm}

\begin{quote}
\begin{lstlisting}[caption={A simplified version of algorithm~\am.},numbers=left,captionpos=b,label=alg:am]
def AM($\alpha, \beta$):
   S = {$(\alpha, \beta)$}
   H = $\emptyset$
   while $(\alpha, \beta)$ = POP(S):
      if $\varepsilon(\alpha) \neq \varepsilon(\beta)$:  return False
      PUSH(H, $(\alpha, \beta)$)
      for $a \in \Sigma$:
         $\alpha^\prime = a^{-1}(\alpha)$
         $\beta^\prime = a^{-1}(\beta)$
         if $(\alpha^\prime, \beta^\prime) \notin H$: PUSH(S,$(\alpha^\prime, \beta^\prime)$)
   return True
\end{lstlisting}
\end{quote}

\subsection{A na\"ive HK algorithm}
\label{subsec:naive}

We now present a na\"ive version of the Algorithm~\ref{alg:hk0}. It
will be useful to prove its correctness and to establish a
relationship to the Antimirov and Mosses' method (\am{}).
Let $A = (Q_1, \Sigma, p_0, \delta_1, F_1)$ and $B = (Q_2, \Sigma,
q_0, \delta_2, F_2)$ be two \dfa{}s, with $|Q_1|=n$ and $|Q_2|=m$, and
$Q_1$ and $Q_2$ disjoint. Consider Algorithm~\ref{alg:hkn}. 
Termination is guaranteed because the number of pairs of states pushed
into $S$ is at most~$mn$ and in each iteration one pair is popped from
$S$.
To prove the correctness we show that in $H$ we collect the pairs of
states of the relation $R$, defined below.

\begin{quote}
\begin{lstlisting}[caption={The algorithm \hkn{}, a na\"ive version of \hk.},captionpos=b,numbers=left,label=alg:hkn]
def HKn(A,B):
   S = $\{(p_0, q_0)\}$
   H = $\emptyset$
   while $(p, q)$ = POP(S):
      PUSH(H,$(p, q)$)
      for $a \in \Sigma$:
         $p^\prime = \delta_1(p, a)$
         $q^\prime = \delta_2(q, a)$
         if $(p^\prime, q^\prime) \notin$ H: PUSH(S,$(p^\prime, q^\prime)$)
   for $(p,q)$ in H:
      if $\const(p)\not= \const(q)$: return False
   return True
\end{lstlisting}
\end{quote}

\begin{lemma}\label{lem:3}
  In Algorithm~\ref{alg:hkn}, for all $(p,q)\in Q_1\times Q_2$,
  $(p,q)\in S$ in a step $k>0$ if and only if $(p,q)\in H$ for some
  step $k' > k$.
\end{lemma}

\begin{definition}
  \label{def:1}
  Let $R$ be defined as follows:
$$ R =\{(p,q)\in Q_1\times Q_2 \mid \exists x \in \Sigma^\star \, :
\delta_1(p_0,x)=p\, \wedge \delta_2(q_0,x)=q\}.$$
\end{definition}

\begin{lemma} 
  \label{lem:4}
  For all $(p,q)\in Q_1\times Q_2$, $(p,q)\in S$ at some step of
  Algorithm~\ref{alg:hkn}, if and only if~$(p,q)\in R$.
\end{lemma}

\begin{lemma}
  \label{lem:5}
  In line $10$, for all $(p,q)\in Q_1\times Q_2$, $(p,q)\in R$ if and
  only if~$(p,q)\!\in\! H$.
\end{lemma}

Considering Lemma~\ref{lem:4} and Lemma~\ref{lem:5}, the following
theorem ensures the correctness of Algorithm~\ref{alg:hkn}.

\begin{theorem}
  \label{th:dfa_equiv}
  $A\sim B$ if and only if for all $(p,q)\in R,\;\const(p)=\const(q)$.
  \begin{proof}
  \emph{
      Suppose, by absurd, that $A$ and $B$ are not equivalent and that
      the condition holds. Then, there exists $w\in\Sigma^\star$ such
      that $\const(\delta(p_0,w))\not = \const(\delta(q_0,w))$. But in
      that case there is a contradiction because $(\delta(p_0,w),
      \delta(q_0,w)) \in R$. On the other hand, if there exists a
      $(p,q)\in R$ such that $\const(p)\not=\const(q)$, obviously $A$
      and $B$ are not equivalent.}
  \end{proof}
\end{theorem}

The relation $R$ can be seen as a relation on $(Q_1\cup Q_2)^2$ which
is reflexive and symmetric. Its transitive closure $R^\star$ is an
equivalence relation.

\begin{lemma}\label{lem:rs}
  $\forall (p,q)\in R, \;\const(p)=\const(q) \text{ if and only if }\;
  \forall (p,q)\in R^\star, \;\const(p)=\const(q).$
\end{lemma}
\begin{corollary}\label{cor:1}
 $A\sim B$ if and only if $\;\forall (p,q)\in R^\star,\;\const(p)=\const(q)$.
\end{corollary}
  The Algorithm \hk{} computes $R^\star$ by starting with the finest
  partition in $Q_1\cup Q_2$ (the identity). And if $A\sim B$,
  $R^\star$ is a right-invariance.
\begin{corollary}\label{cor:hknhk0}
  Algorithm~\ref{alg:hkn} and Algorithm~\ref{alg:hk0} are equivalent.
\end{corollary}

\subsection{Equivalence of the two methods}

The Algorithm~\ref{alg:hkn} can be modified to a \emph{earlier
  refutation} version, as in Algorithm~\ref{alg:hk1}. In order to do
so, we remove lines 10--11, and we insert a line equal to line 7 of
Algorithm~\ref{alg:hk1}, before line 4. It is then obvious that
Algorithm \ref{alg:am} corresponds to Algorithm \ref{alg:hkn} applied
to Brzozowski's automata of two \re, where these \dfas are
incrementally constructed during the algorithm's execution. In
particular, the halting conditions are the same considering the
definition of final states in a Brzozowski's automaton.

\begin{theorem}\label{theor:amhk}
  Algorithm~\ref{alg:am} (\am) corresponds to Algorithm~\ref{alg:hkn} (\hkn) applied to Brzozowski's
  automata of two regular expressions.
\end{theorem}



\vspace{-0.3cm}
\subsection{Improving Algorithm \am with Union-Find}

Considering the Theorem~\ref{theor:amhk} and the
Corollary~\ref{cor:hknhk0}, we can improve the Algorithm~\ref{alg:am}
(\am) for testing the equivalence of two \re{} $\alpha$
and $\beta$, by considering Algorithm~\ref{alg:hk0} applied to the
Brzozowski's automata correspondent to the two \re Instead of using a
stack ($H$) in order to keep an history of the pairs of regular
expressions which have already been tested, we can build the
correspondent equivalence relation~$R^\star$ (as defined for
Lemma~\ref{lem:rs}). Two main changes must be considered:
\begin{itemize}
\item One must ensure that the sets of derivatives of each regular
  expression are disjoint. For that we consider their disjoint sum,
  where derivatives w.\,r.\,t. a word $u$ are represented by tuples
  $(\derivative{u}(\alpha),1)$ and $(\derivative{u}(\beta),2)$,
  respectively.
\item In the UNION-FIND method, the FIND operation needs an equality
  test on the elements of the set. Testing the equality of two \re ---
  even syntactic equality --- is already a computationally expensive
  operation, and tuple comparison will be even slower. On the other
  hand, integer comparison, can be considered to be $O(1)$. As we know
  that each element of the set is unique, we may consider some hash
  function which assures that the probability of collision for these
  elements is extremely low. This allows us to safely use the hash
  values as the elements of the set, and thus, arguments to the FIND
  operation, instead of the \re themselves. This is also a natural
  procedure in the implementations of conversions from \re to
  automata.
\end{itemize}

We call \equivuf{} to the resulting algorithm. The experimental results
are presented on Table \ref{tab:re}, Section \ref{sec:experimental}.

\vspace{-0.2cm}
\subsection{Worst-case complexity analysis}

In Almeida \emph{et al.}
\cite{almeida08_c:_antim_and_mosses_rewrit_system_revis} the algorithm \am was improved by  considering partial
derivatives~\cite{antimirov96:_partial_deriv_regul_expres_finit_autom_const}.  The
resulting algorithm (\equivp) can be seen as the algorithm \hke
applied to the partial derivatives \nfa of a \re We present a lower
bound for the worst-case complexity of this algorithm by exhibiting a
family of \re for which the comparison method can be exponential on the number of alphabetical symbols
$|\alpha|_\Sigma$ of a \re~$\alpha$. We will proceed by showing that
the partial derivatives \nfa $N$ of a \re
$\alpha$
is such that $|N| \in O(|\alpha|_\Sigma)$ and the number of states of
the smallest equivalent \dfa is exponential on $|N|$.

\begin{figure}[ht]
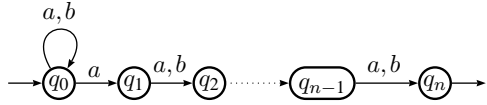

  \begin{center}
    \SmallPicture {
      \VCDraw {
        \begin{VCPicture}{(-2,0)(12,2)}
          \State[q_0]{(0,0)}{A} \State[q_1]{(2,0)}{B}
          \State[q_2]{(4,0)}{C}
          \StateVar[q_{n-1}]{(7,0)}{E}
          \State[q_n]{(10,0)}{F}
          \Initial{A} \Final{F}
          \LoopN[.5]{A}{a,b} \EdgeL{A}{B}{a} \EdgeL{B}{C}{a,b}
          \ChgEdgeLineStyle{dotted}\EdgeL{C}{E}{}\RstEdgeLineStyle 
          \EdgeL{E}{F}{a,b}
        \end{VCPicture}
      }
    }
  \end{center}\vspace{1mm}
  \caption{NFA which has no equivalent DFA with less than $2^n$
    states.}
  \label{fig:exp_dfa}
\end{figure}

Figure~\ref{fig:exp_dfa} presents a classical example of a bad
behaved case of the powerset construction, by Hopcroft \emph{et al.}
\cite{hopcroft00_c:_introd_autom_theor_languag_comput}. Although this
example does not reach the $2^n$ states bound, the smallest equivalent
\dfa has exactly~$2^{n-1}$ states.

\begin{figure}[ht]
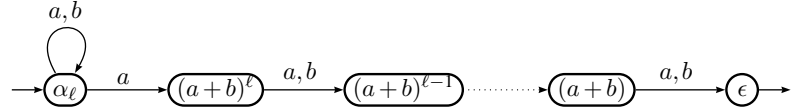

  \begin{center}
    \SmallPicture {
      \VCDraw {
        \begin{VCPicture}{(-2,0)(18,2)}
          \StateVar[\alpha_\ell]{(0,0)}{A} \StateVar[(a+b)^\ell]{(4,0)}{B}
          \StateVar[(a+b)^{\ell-1}]{(9,0)}{C} 
          \StateVar[(a+b)]{(14,0)}{E} \StateVar[\epsilon]{(18,0)}{F}
          \Initial{A} \Final{F}
          \LoopN[.5]{A}{a,b} \EdgeL{A}{B}{a} \EdgeL{B}{C}{a,b}
          \ChgEdgeLineStyle{dotted}\EdgeL{C}{E}{}\RstEdgeLineStyle
          \EdgeL{E}{F}{a,b}
        \end{VCPicture}
      }
    }
  \end{center}\vspace{1mm}
  \caption{\nfa obtained from the \re $\alpha$ using the \am method.}
  \label{fig:pd_nfa}
\end{figure}

Consider the \re family $\alpha_\ell = (a+b)^\star a
(a+b)^\ell$, where $|\alpha_\ell|_\Sigma=3+2\ell=m$. It is easy to see
that the \nfa in Figure \ref{fig:exp_dfa} is obtained directly from
the application of the \am method to $\alpha_\ell$, with the corresponding
partial derivatives presented on Figure \ref{fig:pd_nfa}.

The set of the partial derivatives
$$PD(\alpha_\ell) = \{ \alpha_\ell,(a+b)^\ell, \dots, (a+b), \epsilon \}$$
has $\ell+2=\frac{m+1}{2}$
elements, which corresponds to the size of the obtained \nfa. The
equivalent minimal \dfa has $2^{\ell+1}=2^{\frac{m-1}{2}}$ states.

\section{Experimental results}
\label{sec:experimental}

In this section we present some experimental results of the previously
discussed algorithms applied to \dfas, \nfas, and \re We also include
the same results of the tests using Hopcroft's (\hop{}) and
Brzozowski's (\brz{})
\cite{brzozowski63_c:_canon_regul_expres_and_minim} 
automata
minimization algorithms. The random \dfas were generated using
publicly available
tools\footnote{{\tt http://www.ncc.up.pt/FAdo/node1.html}}\cite{almeida07_c:_enumer_gener_strin_autom_repres}. The \nfas
dataset was obtained with a set of tools described by Almeida \emph{et
  al.} \cite{almeida08:_perfor_of_autom_minim_algor}.
All the algorithms were implemented in the \texttt{Python} programming
language. The tests were executed in the same computer, an
Intel\textsuperscript{\textregistered}
Xeon\textsuperscript{\textregistered} 5140 at 2.33GHz with 4GB of RAM.
\begin{table}[ht]
  \caption{Running times for tests with complete accessible \dfas.}
  \label{tab:icdfa}
  \centering
  \footnotesize
  \begin{tabular}[h]{|l|c|c|c|c|c|c|c|c|c|c|c|c|c|c|c|}
    \hline
    &\multicolumn{6}{|c|}{$n=5$} & \multicolumn{6}{|c|}{$n=50$} \\ 
    \hline
    &\multicolumn{3}{|c|}{$k=2$} & \multicolumn{3}{|c|}{$k=50$} &
    \multicolumn{3}{|c|}{$k=2$} & \multicolumn{3}{|c|}{$k=50$} \\
    \hline
    \bf Alg. &\multicolumn{2}{|c|}{\bf Time (s)} & {\bf Iter.} &
    \multicolumn{2}{|c|}{\bf Time (s)} & {\bf Iter.} &
    \multicolumn{2}{|c|}{\bf Time (s)} & {\bf Iter.} &
    \multicolumn{2}{|c|}{\bf Time (s)} & {\bf Iter.} \\
    \hline
    &Eff. & Total & Avg. & Eff. & Total & Avg. & Eff. & Total & Avg. & Eff. & Total & Avg. \\
    \hline
    \bf Hop	&5.3	&7.3	&-    &85.2	&91.0	&-     &566.8   &572	&- &17749.7 &17787.5	&- \\
    \bf Brz	&25.5	&28.0	&-    &1393.6	&1398.9 &-     &-	&-	&-   &-	   &-	        &- \\
    \bf HK	&2.3	&4.0	&8.9  &25.3	&28.9	&9.0   &23.2	&28.9	&98.9 &317.5   &341.6	&99.0 \\
    \bf HKe	&0.9	&2.1	&2.4  &5.4	&10.5	&2.4   &1.4	&5.9	&2.6 &14.3	   &34.9	&3.4 \\
    \bf HKs     &0.6	&1.3	&2.4  &2.8	&4.6	&2.4   &0.8	&2.0	&2.7 &9.1	   &21.3	&3.4 \\
    \bf HKn	&0.7	&2.2	&3.0  &51.5     &56.2   &29.7  &1.3	&6.8	&3.7 &29.4	   &51.7	&15.4 \\
    \hline
  \end{tabular}
\end{table}
 Table \ref{tab:icdfa} shows the results of
experimental tests with $10.000$ pairs of complete \icdfa{}s. Due to
space constraints, we only present the results for automata with $n
\in \{5, 50\}$ states over an alphabet of $k \in \{2, 50\}$
symbols. Clearly, the methods which do not rely in minimisation
processes are \emph{a lot} faster.  Below (Eff.) appears the
\emph{effective} time spent by the algorithm itself while below
(Total) we show the \emph{total} time spent, including overheads,
such as making a \dfa complete, initializing auxiliary data
structures, etc. All times are expressed in seconds, and the
algorithms that were not finished after 10 hours are accordingly
signaled. The algorithm \brz{} is by far the slowest. The algorithm
\hop, although faster, is still several orders of magnitude slower
than any of the algorithms of the previous sections. We also present
the average number of iterations (Iter.) used by each of the versions
of algorithm \hk, per pair of automata. Clearly, the refutation
process is an advantage. \hkn{} running times show that a linear set
merging algorithm (such as UNION-FIND) is by far a better choice than
a simple history (set) with pairs of states. \hks{} is a version of
\hke{} which uses the automata string representation proposed by Almeida \emph{et al.}~\cite{almeida07_c:_enumer_gener_strin_autom_repres,reis05_c:_repres_finit_dcfs}. The simplicity of
the representation seemed to be quite suitable for this algorithm, and
actually cut down both running times to roughly half. This is an
example of the impact that a good data structure may have on the
overall performance of this algorithm.

\begin{table}[ht]
  \caption{Running times for tests with 10.000 random \nfas.}
  \label{tab:nfa}
  \centering
  \footnotesize
  \begin{tabular}[h]{|l|c|c|c|c|c|c|c|c|c|c|c|c|c|c|c|}
    \hline
    &\multicolumn{6}{|c|}{$n=5$} & \multicolumn{6}{|c|}{$n=50$} \\ 
    \hline
    &\multicolumn{3}{|c|}{$k=2$} & \multicolumn{3}{|c|}{$k=20$} &
    \multicolumn{3}{|c|}{$k=2$} & \multicolumn{3}{|c|}{$k=20$} \\
    \hline
    \bf Alg. &\multicolumn{2}{|c|}{\bf Time (s)} & {\bf Iter.} &
    \multicolumn{2}{|c|}{\bf Time (s)} & {\bf Iter.} &
    \multicolumn{2}{|c|}{\bf Time (s)} & {\bf Iter.} &
    \multicolumn{2}{|c|}{\bf Time (s)} & {\bf Iter.} \\
    \hline
    &Eff. & Total & Avg. & Eff. & Total & Avg. & Eff. & Total & Avg. & Eff. & Total & Avg. \\
    \hline
    \multicolumn{13}{|c|}{Transition Density $d=0.1$} \\
    \hline
    \bf Hop &10.3	&12.5	&-	&1994.7	&2003.2	&-	&660.1	&672.9	&-	&-	&-	&-	\\
    \bf Brz &8.4	&10.6	&-	&866.6	&876.2	&-	&264.5	&278.4	&-	&-	&-	&-	\\
    \bf HKe &0.8	&2.9	&2.2	&8.4	&19	&4	&24.4	&37.8	&10.2	&-	&-	&-	\\
    \hline
    \multicolumn{13}{|c|}{Transition Density $d=0.5$} \\
    \hline
    \bf Hop &17.9	&19.8	&-	&2759.4	&2767.5	&-	&538.7	&572.6	&-	&-	&-	&-	\\
    \bf Brz &14.4	&16	&-	&2189.3	&2191.6	&-	&614.9	&655.7	&-	&-	&-	&-	\\
    \bf HKe &2.6	&4.3	&4.9	&36.3	&47.3	&10.3	&6.8	&48.9	&2.5	&294.6	&702.3	&11.5	\\
    \hline
    \multicolumn{13}{|c|}{Transition Density $d=0.8$} \\
    \hline
    \bf Hop &12.5	&14.3	&-	&376.9	&385.5	&-	&1087.3	&1134.2	&-	&-	&-	&-	\\
    \bf Brz &14	&15.8	&-	&177	&179.6	&-	&957.5	&1014.3	&-	&-	&-	&-	\\
    \bf HKe &1.4	&3.2	&2.7	&39	&49.9	&10.7	&7.3	&64.8	&2.5	&440.5	&986.6	&11.5	\\
    \hline
  \end{tabular}
\end{table}

Table \ref{tab:nfa} shows the results of applying the same set of
algorithms to \nfas. The testing conditions and notation are as
before, adding only the \emph{transition density}~$d$ as a new
variable, which we define as the ratio of the number of transitions
over the total number of possible transitions ($kn^2$). Although it is
clear that \hke is faster, by at least one order of magnitude, than
any of the other algorithms, the peculiar behaviour of this algorithm
with different transition densities is not easy to
explain. Considering the simplest example of 5 states and 2 symbols,
the dataset with a transition density $d=0.5$ took roughly twice as
long as those with $d \in \{ 0.1, 0.8\}$. On the other extreme, making
$n=50$ and $k=2$, the hardest instance was $d=0.1$, with the cases
where $d \in \{0.5, 0.8\}$ present similar running times almost five
times faster. In our largest test, with $n=50$ and $k=20$, neither
\hop nor \brz finished within the imposed time limit. Again, $d=0.1$
was the hardest instance for \hke, which also did not finish within
the time limit, although the cases where $d \in \{0.5, 0.8\}$ present
similar running times.

\begin{table}[ht]
  \caption{Running times (seconds) for tests with 10.000 random \re}
  \label{tab:re}
  \centering
  \footnotesize
  \begin{tabular}[h]{|l|c|c|c|c|c|c|c|c|}
    \hline
    &\textbf{Size/Alg.}	&\textbf{Hop} &\textbf{Brz} &\textbf{AM} &\textbf{Equiv} &\textbf{EquivP} &\textbf{HKe} &\textbf{EquivUF} \\
    \hline
    $k=2$&10	&21.025	&19.06	&26.27	&7.78	&5.512	&7.27 &5.10 \\
    &50	&319.56	&217.54	&297.23	&36.13	&28.05	&64.12 &28.69 \\
    &75	&1043.13	&600.14	&434.89	&35.79	&23.46	&139.12 &60.09 \\
    &100	&7019.61	&1729.05	&970.36	&60.76	&48.29	&183.55 &124.00 \\
    \hline
    $k=5$&10	&42.06	&25.99	&32.73	&9.96	&7.25	&8.69 &6.48 \\
    &50	&518.16	&156.28	&205.41	&33.75	&26.84	&67.7 &21.53 \\
    &75	&943.65	&267.12	&292.78	&35.09	&25.17 &161.84 &28.61 \\
    &100	&1974.01	&386.72	&567.39	&54.79	&45.41	&196.13 &37.02 \\
    \hline
    $k=10$&10	&61.60	&31.04	&38.27	&10.87	&8.39	&9.26 &7.47 \\
    &50	&1138.28	&198.97	&184.93	&34.93	&28.95	&72.95 &22.60 \\
    &75	&2012.43	&320.37	&271.14	&35.77	&26.92	&195.88 &30.61 \\
    &100	&4689.38	&460.84	&424.67	&52.97	&44.58	&194.01 &39.23 \\
    \hline
  \end{tabular}
\end{table}

Table \ref{tab:re} presents the running times of the application of
\hke to \re{} and their comparison with the algorithms presented by
Almeida \emph{et al.}
\cite{almeida08_c:_antim_and_mosses_rewrit_system_revis}, where
\textbf{equiv} and \equivp are the functional variants of the original
\am algorithm. \equivuf is the UNION-FIND improved version of \equivp.
Although the results indicate that \hke{} is not as fast as the direct
comparison methods presented in the cited paper, it is clearly faster
than any minimisation process. The improvements of \equivuf over
\equivp are not significant (it is actually considerably slower for
\re of length 100 with 2 symbols). We suspect that this is related to
some optimizations applied by the \texttt{Python} interpreter. We
state this based on the fact that when both algorithms are executed
using a profiler, \equivuf is almost twice faster than \equivp on most
tests.

We have no reason to believe that similar tests with different
implementations of these algorithms would produce significantly
different ordering of its running times from the one here
presented. However, it is important to keep in mind, that these are
experimental tests that greatly depend on the hardware, data
structures, and several implementation details (some of which, such as
compiler optimizations, we do not utterly control).

\section{Conclusions}
As minimality or equivalence for (finite) transition systems is in
general intractable, right-invariant relations (bisimulations) have
been extensively studied for nondeterministic variants of these
systems. When considering deterministic systems, however, those
relations provide non-trivial improvements. We presented several
variants of a method by Hopcroft and Karp for the comparison of \dfas
which does not use automata minimization. By placing a refutation
condition earlier in the algorithm we may achieve better running times
in the average case. This is sustained by the experimental results
presented in the paper. We extended this algorithm to handle
\nfas. Using Brzozowski's automata, we showed that a modified version
of Antimirov and Mosses' method translates directly to Hopcroft and
Karp's algorithm.

\bibliographystyle{eptcs}
\bibliography{almeida}
\end{document}